\documentclass[aps,prl,twocolumn,showpacs,superscriptaddress]{revtex4}
\usepackage{graphicx,color}
\usepackage{amsmath}
\usepackage{amssymb}
\usepackage{bm}

\newcommand{\ket}[1]{|{#1}\rangle}
\newcommand{\bra}[1]{\langle{#1}|}

\begin{document}
\title{Entanglement of Electrons Field-Emitted from a Superconductor}

\author{Kazuya Yuasa}
\affiliation{Waseda Institute for Advanced Study, Waseda University, Tokyo 169-8050, Japan}

\author{Paolo Facchi}
\affiliation{Dipartimento di Matematica, Universit\`a di Bari, I-70125 Bari, Italy}
\affiliation{Istituto Nazionale di Fisica Nucleare, Sezione di Bari, I-70126 Bari, Italy}

\author{Rosario Fazio}
\affiliation{NEST-CNR-INFM \& Scuola Normale Superiore, piazza dei Cavalieri 7, I-56126 Pisa, Italy}
\affiliation{International School for Advanced Studies (SISSA), via Beirut 2-4, I-34014 Trieste, Italy}

\author{Hiromichi~Nakazato}
\affiliation{Department of Physics, Waseda University, Tokyo 169-8555, Japan}

\author{Ichiro Ohba}
\affiliation{Department of Physics, Waseda University, Tokyo 169-8555, Japan}

\author{Saverio Pascazio}
\affiliation{Dipartimento di Fisica, Universit\`a di Bari, I-70126 Bari, Italy}
\affiliation{Istituto Nazionale di Fisica Nucleare, Sezione di Bari, I-70126 Bari, Italy}

\author{Shuichi Tasaki}
\affiliation{Department of Applied Physics and Advanced Institute for Complex Systems,
Waseda University, Tokyo 169-8555, Japan}

\date[]{April 22, 2009}

\begin{abstract}
Under appropriate circumstances the electrons emitted from a superconducting tip can be entangled.
We analyze these nonlocal correlations by studying  the coincidences of the field-emitted electrons and 
show that electrons emitted in opposite directions violate Bell's inequality. 
We scrutinize the interplay between the bosonic nature of Cooper pairs and the fermionic nature of electrons. 
We further discuss the feasibility of our analysis in the light of present experimental capabilities.
\end{abstract}

\maketitle

Entanglement, at the heart of the foundations of quantum mechanics, has gained renewed attention with the birth of quantum information science \cite{nielsen}.
Here it is considered  to be a precious resource, as it is believed to be the key ingredient for the increased efficiency of quantum protocols  as compared with their classical counterparts.  
Finding sources of entangled particles is therefore of paramount importance. 
In quantum optics this is well known and routinely used.  
For example, in parametric down conversion \cite{mandel} a pump laser beam incident on a non-linear crystal leads to the generation of entangled photon pairs. 
In electronics, the field is much younger, but there are already a number of very interesting proposals to generate entangled  electron states (see the reviews \cite{ref:reviews}). 
In this Rapid Communication, we propose an \textit{entangled electron} source \textit{in vacuum}, on the basis of a thorough analysis of the entanglement and the correlations of the electrons field-emitted from a superconductor.

When a bias voltage is applied to a sharp piece of material, a strong electrostatic field is realized at the tip, causing electron emission into vacuum. 
The ground state of superconductor is  a fairly controllable macroscopic quantum state and provides a coherent and monochromatic electron beam via field emission from a 
superconducting tip, as experimentally shown in \cite{ref:OshimaNb}.
Our analysis will show that electron coincidences in field emission can reveal electron non-local  
correlations due to pairing in the superconducting tip. 
Moreover, we shall see that, by orienting the detectors in opposite directions, one can optimize the fraction of entangled electrons in order to perform a test of Bell's inequality.

Field emission thus enables one to study Bell's inequality on electrons in vacuum by means of 
correlation measurements. Signatures of quantum statistics and of correlation can be
unambiguously detected by coincident  measurements. 
After the seminal result by Hanbury Brown and Twiss \cite{ref:HBT}, bunching and
antibunching of bosons and fermions have been
measured in a series of important experiments \cite{ref:Bunching-Photon,ref:Bunching-Atom,ref:Antibunching-Atom,ref:Antibunching-Solid,ref:Antibunching-Kiesel,ref:Antibunching-Neutron}.
Advanced technology in single-electron detection has made possible
the observation of antibunching in field emission \cite{ref:Antibunching-Kiesel}. 
An experiment with a superconducting tip, similar to \cite{ref:OshimaNb}, is what is needed
to test our proposal.

We will show that the spectrum of emitted electronics from a superconductor displays a 
remarkable  interplay between positive correlations (bunching), due to the bosonic 
nature of the Cooper pairs, and negative correlations (antibunching) due to the fermionic 
nature of the electrons that make up the Cooper pairs.  
An electron pair exhibiting positive correlation is in a Bell state. 
The presence of a bunching-like behavior in quantum transport through
multiterminal superconductor-normal metal structures has already been pointed out 
through an analysis of current noise \cite{ref:blanter,ref:PositiveCorr}. 
Superconductors were 
also shown to be a source of entangled electrons, whose generation and
detection in hybrid conductors has been
discussed in a number of articles \cite{ref:supent}.
In particular, we draw attention to the work of Prada and Sols in \cite{ref:supent}, where
the angular distribution of the emitted
electrons from a superconductor has been discussed.

\begin{figure}[t]
\includegraphics[width=0.44\textwidth]{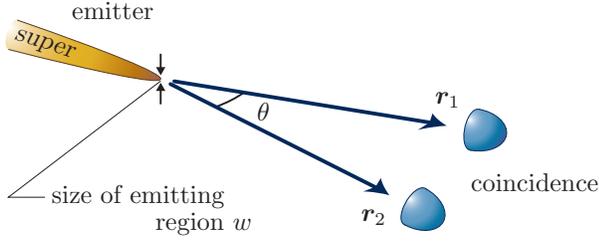}
\caption{(Color online) Field emission of electrons and coincident detections.
The size of the emitting region at the tip is $w$ and two of the emitted electrons are detected at $\bm{r}_1$ and $\bm{r}_2$.}
\label{fig:Setup}
\end{figure}

Our setup is sketched in Fig.\ \ref{fig:Setup}. The Hamiltonian in
3D space reads $H=H_S+H_V+H_T$, with  ($\hbar=1$)
\begin{align}
&H_S
=\sum_{s=\uparrow,\downarrow}\int d^3\bm{k}\,\omega_k
\alpha_{\bm{k}s}^\dag\alpha_{\bm{k}s},\quad
\omega_k=\sqrt{\varepsilon_k^2+|\Delta|^2},
\label{eqn:Hamiltonian}
\displaybreak[0]
\\
&H_V
=\sum_{s=\uparrow,\downarrow}\int d^3\bm{p}\,\varepsilon_p
c_{\bm{p}s}^\dag c_{\bm{p}s},\quad\ \ %
\varepsilon_p=\frac{p^2}{2m}-\mu,
\displaybreak[0]
\\
&H_T
=\sum_{s=\uparrow,\downarrow}
\int d^3\bm{p}\int d^3\bm{k}\,(
T_{\bm{p}\bm{k}}c_{\bm{p}s}^\dag a_{\bm{k}s}
+T_{\bm{p}\bm{k}}^*a_{\bm{k}s}^\dag c_{\bm{p}s}
),
\label{eqn:HT}
\end{align}
where $H_S$ is the Hamiltonian of the superconducting emitter,
$\alpha_{\bm{k}s}$ the fermionic operators of the quasiparticles,
$H_V$ describes the propagation of the electrons in vacuum,
$c_{\bm{p}s}$ are fermionic operators, and the (interaction)
Hamiltonian $H_T$ describes the emission of the electrons
from the superconductor into
vacuum \cite{ref:Tunneling,ref:FEAndreev,ref:LateralEffects}. 
The energy of an electron in vacuum, $\varepsilon_p$, as well as that of
a quasiparticle in the superconductor,
$\omega_k$, are both measured from the Fermi level of the superconductor, $\mu$.
$\Delta$ is the gap of the superconductor and the quasiparticle operators
$\alpha_{\bm{k}s}$ are related to the electron operators $a_{\bm{k}s}$ by a Bogoliubov 
transformation \cite{Tinkham}.
The Coulomb interaction among the emitted electrons can be safely neglected since it 
becomes relevant at much larger current densities, as compared to those typical of the experiments relevant for the present work \cite{note:Coulomb}.

For the tunneling matrix elements we take \cite{ref:LateralEffects}
\begin{gather}
T_{\bm{p}\bm{k}}
= h(\bm{p})g (\bm{p}-\bm{k}),\\
g(\bm{p})=(2\pi)^{-3} e^{-p^2w^2/2},\quad h(\bm{p})=(p/m)^{1/2}
e^{\varepsilon_p/2E_C},
\label{eqn:ff}
\end{gather}
where $g(\bm{p})$ characterizes the emitting region of size $w$ and
$h(\bm{p})$ the tunneling probability, that decays
exponentially as the incident energy
decreases \cite{note:NormalEnergy}, $E_C$ being an energy
scale that characterizes the energy dependence of the tunneling
matrix elements.

The emission process is dynamically described in 3D space with the Hamiltonian
(\ref{eqn:Hamiltonian})--(\ref{eqn:HT}) and after a transient period it reaches a
nonequilibrium steady state \cite{ref:Ruelle}, with a stationary beam of electrons
emitted from the superconductor. Electron correlations are present in this
beam \cite{ref:LateralEffects}. When the detectors do not resolve the spin states of the electrons,
the probability of two joint detections, at $(\bm{r}_1,t_1)=(1)$ and $(\bm{r}_2,t_2)=(2)$, with $t_2\ge t_1$, is proportional to
\begin{align}
\rho^{(2)}(2;1)
&=\sum_{s_1,s_2=\uparrow,\downarrow}
\langle\psi_{s_1}^\dag(1)\psi_{s_2}^\dag(2)
\psi_{s_2}(2)\psi_{s_1}(1)\rangle\nonumber\\
&=4\gamma(2;2)\gamma(1;1)
-2|\gamma(2;1)|^2
+2|\chi(2;1)|^2,
\label{eqn:TwoDistri}
\end{align}
where
$\psi_s(\bm{r},t)$
is the field operator of the electrons in vacuum, 
\begin{align}
\gamma(2;1)
&=\langle\psi_\uparrow^\dag(2)\psi_\uparrow(1)\rangle
=\langle\psi_\downarrow^\dag(2)\psi_\downarrow(1)\rangle,\\
\chi(2;1)
&=\langle\psi_\uparrow(2)\psi_\downarrow(1)\rangle
=-\langle\psi_\downarrow(2)\psi_\uparrow(1)\rangle.
\label{eqn:ChiDef}
\end{align}
The correlation function $\gamma$ describes the state of single
electrons, and, in particular, the (spin-summed) one-particle
distribution of the emitted electrons is given by
$\rho^{(1)}(\bm{r},t)=2\gamma(\bm{r},t;\bm{r},t)$. The
correlation function $\chi$ describes the emission of \textit{pairs}
of electrons (with opposite spins). A second-order calculation for
the coincident detections at $t_1=t_2$ yields
\begin{widetext}
\begin{equation}
\chi(2;1)
=\int d^3\bm{k}\,u_kv_k
\int\frac{d^3\bm{p}_1}{\sqrt{(2\pi)^3}}
\int\frac{d^3\bm{p}_2}{\sqrt{(2\pi)^3}}
\frac{T_{\bm{p}_1\bm{k}}T_{\bm{p}_2(-\bm{k})}
}{\varepsilon_{p_1}+\varepsilon_{p_2}-i0^+}
\left(
\frac{1}{\varepsilon_{p_1}+\omega_k-i0^+}
+\frac{1}{\varepsilon_{p_2}+\omega_k-i0^+}
\right)
e^{i\bm{p}_1\cdot\bm{r}_1}e^{i\bm{p}_2\cdot\bm{r}_2},
\label{eqn:Andreev}
\end{equation}
\end{widetext}
where $u_k$ and $v_k$ are the Bogoliubov amplitudes.
Notice that, since $u_kv_k=\Delta/2\omega_k$, $\chi$
is proportional to the gap parameter $\Delta$ and vanishes when the emitter is in its normal state.
At zero temperature, there is no contribution from the quasiparticle excitations and Eq.\ (\ref{eqn:Andreev})
is due to Andreev processes.
In the absence of this contribution, the second term in (\ref{eqn:TwoDistri}) reduces the coincidence probability of finding two electrons close to each other within a small time delay, exhibiting antibunching.
The pair correlation $\chi$, on the other hand, enhances such coincidence probability.
This is relevant for the occurrence of positive correlations \cite{ref:blanter,ref:PositiveCorr}.
We now analyze these effects in greater details.

\begin{figure}[b]
\includegraphics[width=0.45\textwidth]{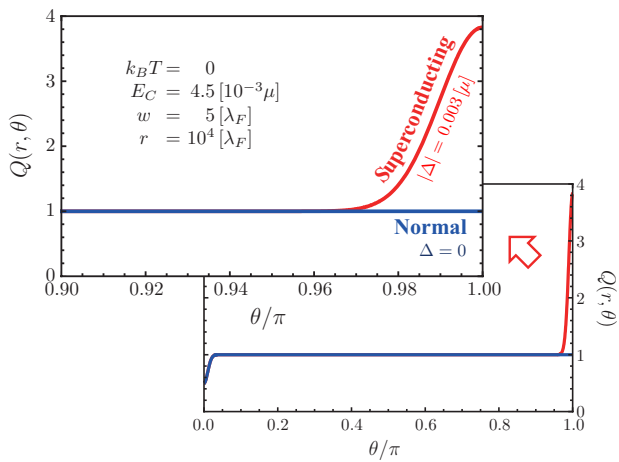}
\caption{(Color online) Normalized coincidences $Q(r,\theta)$
vs $\theta$, for normal $\Delta=0$ and superconducting
$\Delta\neq0$ emitters. Observe the large bunching peak at $\theta \simeq \pi$.}
\label{fig:CorrelationAngle}
\end{figure}
The normalized coincidence
\begin{equation}
Q(r,\theta)=\frac{\rho^{(2)}(2;1)}{\rho^{(1)}(2)\, \rho^{(1)}(1)}
\end{equation}
at $t_1=t_2$, when the detectors are at the same distances
$r_1=r_2=r$ from the tip, is plotted in Fig.\
\ref{fig:CorrelationAngle} as a function of the angle $\theta$
between $ \bm{r}_1$ and $ \bm{r}_2$, for normal and superconducting
emitters. Here, $k_F=\sqrt{2m\mu}=2\pi/\lambda_F$. The effects
of superconductivity are manifest: a bunching peak appears at
$\theta\simeq\pi$. Its origin is clear from the expression of the
Andreev process (\ref{eqn:Andreev}) ($\gamma$ is negligibly small at
$\theta\simeq\pi$).
This shows that electrons with opposite momenta $\bm{k}$ and $-\bm{k}$ are emitted in pair through a virtual process
and propagate with momenta $\bm{p}_1$ and $\bm{p}_2$ in vacuum, in approximately opposite directions (with unavoidable
diffraction effects governed by the size of the emitting region $w$).
The couple $\bm{k}$ and $-\bm{k}$ reflects the Cooper-pair correlation in the emitter.
Notice that the integrand of
$\chi$ in (\ref{eqn:Andreev}) is symmetric under the exchange
$\bm{k}\leftrightarrow-\bm{k}$.
This is because the Cooper pair is in a singlet spin state.
This symmetry yields bunching, which is observed in opposite directions.

Bunching is therefore a signature of excess singlet pairs, when the emissions take place in opposite directions.
It is then of great interest to discuss the nonlocal aspects of the phenomenon \cite{ref:supent,ref:reviews}.
The spin state $\varrho$ of the pair of emitted electrons
is
\begin{align}
\varrho\propto{}&
\Bigl(
\gamma(2;2)\gamma(1;1)
-|\gamma(2;1)|^2
\Bigr)
\,\openone
\nonumber\displaybreak[0]\\
&{}+2\,\Bigl(
|\gamma(2;1)|^2+|\chi(2;1)|^2
\Bigr)\,\ket{\Psi^-}\bra{\Psi^-},
\label{eqn:ExtractedState}
\end{align}
where
$\ket{\Psi^-}=(\ket{\uparrow\downarrow}-\ket{\downarrow\uparrow})/\sqrt{2}$
is the singlet state [the normalization factor is given by the
two-particle distribution $\rho^{(2)}(2;1)$ in
(\ref{eqn:TwoDistri})]. Therefore, in general the entanglement of
the singlet component is  masked by the background.

The degree of entanglement is related to the height $\delta Q =
Q(r,\theta)-1$ of the bunching peak at $\theta=\pi$, which, for
$k_Fr\gg1$, $\mu\gg E_C$, and $r/k_Fw^2\gg1$, is given by
\begin{multline}
\delta Q
\sim \frac{\pi^2}{32 K_1^2(|\Delta|/E_C)}
\biggl|
H_0^{(2)}\!\left(
\frac{i w^2}{\pi^2\xi^2}-\frac{r}{2\pi^2 k_F \xi^2}\right)
\\
{}
-\frac{4\Lambda e^{i r/2\pi^2k_F\xi^2} }{\pi\sqrt{i
r/k_Fw^2}}
\biggr|^2,
\label{eqn:Peak}
\end{multline}
where $K_\nu(z)$ is the modified Bessel function of the second kind, $H_\nu^{(2)}(z)$ is the Hankel function of the second kind, 
$\xi= k_F/\pi m|\Delta|$ is Pippard's length, characterizing the
correlation length of the superconductor,
and $\Lambda$ is a smooth bounded function of  $w$, such that $\Lambda\simeq1$ for $w\gtrsim\lambda_F$.
The higher the bunching peak, the larger the entanglement, and eventually
Bell's inequality can be violated. Notice that the electrons of
each pair are emitted in opposite directions and one need not argue
how to separate them.

\begin{figure}[b]
\includegraphics[width=0.48\textwidth]{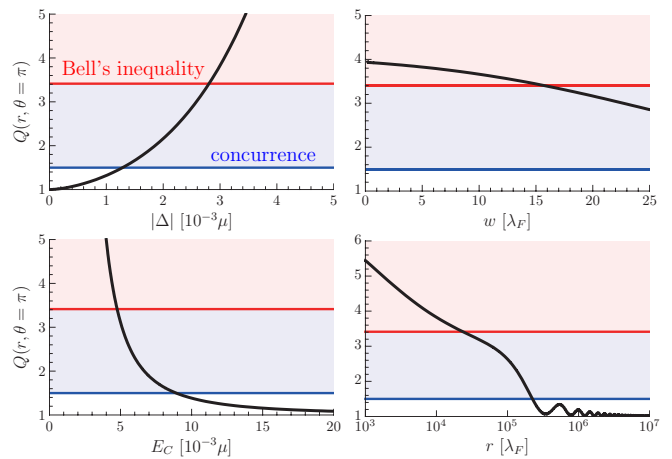}
\caption{(Color online) Peak value of $Q(r,\theta)$
at $\theta=\pi$.
The line at $Q=3/2$ indicates the entanglement threshold (above which the pair of electrons is entangled) and the
one at $Q=\sqrt{2}/(\sqrt{2}-1)\simeq3.41$ is that for the violation
of Bell's inequality (above which Bell's inequality is violated). The parameters are the same as in Fig.\
\ref{fig:CorrelationAngle}, and $\xi\simeq 33.8\,\lambda_F$. }
\label{fig:Peak}
\end{figure}
Figure \ref{fig:Peak} displays the behavior of the bunching peak, 
by scrutinizing the role played by 
the parameters describing the system. Clearly, the value of $\Delta$ is very significant  for entanglement, as
the effects of superconductivity are enhanced. 
By increasing $|\Delta|$, the gap
becomes wider, and
entanglement is enhanced. The parameter $E_C$ appearing in
(\ref{eqn:ff}) works like a filter: by decreasing $E_C$, the
contribution of single-particle emission is suppressed, the
background is reduced, pair emissions become dominant and
entanglement is enhanced.

The effects of $w$ on entanglement are interesting to discuss.
Electron pairs emitted from a smaller
region bunch better and are more entangled. If the emitting region
is larger, there is less guarantee that coincidence electrons
originate from a common Cooper pair, and as a consequence
entanglement is reduced. This explains the role of the ratio between
the size of the emitting region and the extension of a Cooper pair
$w/\xi$, appearing in the formula for the bunching peak
(\ref{eqn:Peak}) and governing the entanglement of the emitted
pairs.

Finally, let us focus on the effects of propagation.
A smaller value of $r$ yields more entanglement. This is
because the wave packets of the emitted electrons spread as they
propagate. Even if two electrons are detected at the same distance
in opposite directions, this does not ensure that the two electrons
originate from a common Cooper pair: there is an ambiguity to the
extent of the spreads of the wave packets. Due to free-space
propagation, the uncertainty at time $t\sim mr/k_F$ is
$\lesssim\sqrt{t/m}$ and this value should be smaller than $\xi$ for
the two electrons to bunch. The bunching peak (\ref{eqn:Peak})
actually decays like $\sim k_F\xi^2/r$ for $r\gg k_F\xi^2$ (with
oscillation), but the length scale $k_F\xi^2$
is much longer than the extension of a Cooper pair $\xi$, and the
slow decay $r^{-1}$ reflects the divergence in the quasiparticle
spectrum.
The oscillations  of $\delta Q$ (below the
entanglement threshold) shown in the last panel of  Fig.\
\ref{fig:Peak} for large values of $r$,  are due to
the asymptotic behavior of the Hankel function $H_0^{(2)}$.

It is important to check to which extent our results are robust  in
a non-ideal situation. To this end we analyzed both  static
fluctuations of the diameter $w$ and the position $\bm{r}_0$ of the emitting tip.
In particular, fluctuations are important only when they become of
order of $\xi$. Moreover, one can show that the angular dependence
of the peak is $\propto \exp\{-8k^2_F w^2
\sin^2[(\pi-\theta)/4]\}$.
Therefore, the effect should be visible as far as $k_F  w\,\delta\theta\lesssim1$, where $\delta
\theta$ is the angle deviation from $\pi$ in the emission of the two
correlated electrons due to  local imperfections of the tip. This
implies a maximum tolerable value of the roughness of the order of
$1/\delta k = 1/k_F\,\delta\theta \simeq w$.

In conclusion we have shown that field emission from a superconducting tip can provide a source of entangled electrons in vacuum.  Besides being of great importance for 
the generation of entanglement in electronics, we believe that a test of Bell's inequality 
on field-emitted electrons is of interest in itself. Moreover, this would be a remarkable example 
in which the interplay between the bosonic nature of Cooper pairs and the fermionic nature of 
electrons is brought to light.  Although all the ingredients to experimentally observe our predictions 
are already available, our analysis shows that  stringent requirements
should be met, as suggested by Fig.\ \ref{fig:Peak}. A large energy resolution and a tip material with a large value of the gap are certainly desirable. Also, energy selection close to the Fermi
level would enhance correlations.

We thank B. Cho, C. Oshima, S. Kawabata, and F. Taddei for discussions.
This work is supported by the bilateral Italian-Japanese Projects II04C1AF4E
of  MUR, Italy,
by the Joint Italian-Japanese Laboratory
of  MAE, Italy,
by the EU
through the Integrated Project EuroSQIP, by the Grant for The 21st Century COE Program
at Waseda University,
the ``Academic Frontier'' Project at Waseda University,
and a Special Coordination Fund for Promoting Science and
Technology from MEXT, Japan,
and by the Grants-in-Aid for Scientific
Research (C)
from JSPS, Japan.


\begin{thebibliography}{10}
\bibitem{note:Coulomb}
Current densities in \cite{ref:OshimaNb} were three or four orders of magnitude lower than the threshold above which the Coulomb effect would become important.  
Even the source current,  $1.5\,\mu\text{A}$, used to measure the  antibunching correlations \cite{ref:Antibunching-Kiesel},  ten to hundred times larger than the source current of \cite{ref:OshimaNb}, is still small enough to neglect Coulomb effects.
\bibitem{note:NormalEnergy}
The tunneling probability should not be a function of $\varepsilon_p$ but rather of the energy associated with the normal motion to the potential surface. In the following discussion, however, we shall look at far field $k_Fr\gg1$, for which this setting is valid.
\end{thebibliography}
\end{document}